\documentclass[twocolumn,prX,superscriptaddress]{revtex4} 
\usepackage[draft]{hyperref}
\usepackage{amsmath}
\usepackage{color}
\usepackage[ansinew]{inputenc}
\usepackage{float}
\usepackage{graphics}

\begin{document}

\title{Observation of continuous-wave squeezed light at 1550\,nm}
\author{Moritz Mehmet}
\affiliation{Max-Planck-Institut f\"ur Gravitationsphysik (Albert-Einstein-Institut) and Institut f\"ur Gravitationsphysik\\ der Leibniz Universit\"at Hannover, Callinstr. 38, 30167 Hannover, Germany}
\affiliation{Centre for Quantum Engineering and Space-Time Research - QUEST, Leibniz
Universit\"at Hannover,\\ Welfengarten 1, 30167 Hannover, Germany}

\author{Sebastian Steinlechner}
\author{Tobias Eberle}
\author{Henning Vahlbruch}
\author{Andr\'e Th\"uring}
\author{Karsten Danzmann}
\author{Roman Schnabel}

\affiliation{Max-Planck-Institut f\"ur Gravitationsphysik (Albert-Einstein-Institut) and Institut f\"ur Gravitationsphysik\\ der Leibniz Universit\"at Hannover, Callinstr. 38, 30167 Hannover, Germany}

\begin{abstract}
{We report on the generation of continuous-wave squeezed vacuum states of light at the telecommunication
wavelength of 1550\,nm. The squeezed vacuum states were produced by type\,I optical parametric amplification
(OPA) in a standing-wave cavity built around a periodically poled potassium titanyl phosphate (PPKTP)
crystal. A non-classical noise reduction of 5.3\,dB below the shot noise was observed by means of balanced
homodyne detection.}
\end{abstract}
\maketitle

%
\noindent
Squeezed states of light were proposed to improve the sensitivity of laser interferometers for the detection
of gravitational waves (GW) \cite{Cav81}, and to establish quantum communication channels \cite{YHa86}, e.g.  for
quantum key distribution~\cite{Ralph99,Hillery00}. For any application of squeezed states of light, a
low decoherence level is required, i.e. optical loss and thermally driven noise sources need to be minimized.
In this respect the laser wavelength of 1550\,nm has emerged as a very interesting topic. Firstly, at this wavelength conventional
silica based telecom glass fibers show low optical loss and can be used for the transmission of squeezed
states. Losses of as low as 0.2\,dB/km were already measured in the late 70's  \cite{Miya79}, and ultra low
loss (ULL) fibers with an attenuation of 0{.}17-0{.}18\,dB/km are commercially available today~\cite{Li08}.
Secondly, at this wavelength, crystalline silicon constitutes an excellent test mass material for
interferometric applications with low optical loss and high mechanical quality \cite{McGuigan}.

GW detectors require the generation of squeezed states in a single spatio-temporal mode of continuous-wave light, whereas quantum channels can also be established in the pulsed laser regime. In the past years, squeezed states at wavelengths beyond 1.5\,$\mu$m were mainly generated in the latter regime. Noise powers of 6.8\,dB below vacuum noise at 1.5$\mu$m\cite{Dong08}, 3.2\,dB at 1.535\,$\mu$m\cite{ETZH07}, and 1.7 dB at 1.55\,$\mu$m\cite{NSHMYG02} were observed. Very recently, continuous-wave squeezed vacuum states at 1560\,nm were generated by an optical parametric oscillator based on periodically poled LiNbO$_3$ (PPLN), and a nonclassical noise suppression of 2.3\,dB was observed \cite{Feng08}. 

Here, we report on the generation of continuous-wave squeezed vacuum states at a wavelength of 1550\,nm
based on periodically poled potassium titanyl phosphate (PPKTP). Squeezing of 5.3\,dB was observed by
balanced homodyne detection. The visibility of the mode-matching between the squeezed field and a
spatially filtered local oscillator beam was measured to be 99\;\%, thereby proving high spatial mode quality of the squeezed states.

\begin{figure}
\center
\includegraphics{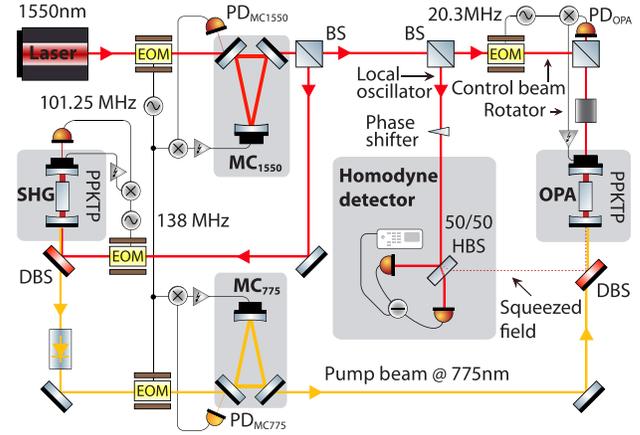}
 \caption{Schematic of the setup. After being sent through a mode cleaner (MC) cavity, one part of the light is used as a control beam for the OPA and the local oscillator for balanced homodyne detection. The other part is frequency doubled in a SHG cavity to provide the 775\,nm field to pump the OPA. The squeezed field leaves the OPA in the counter direction to the pump, and is measured with the homodyne detector.  PBS: polarizing beam splitter; DBS: dichroic beam splitter; HBS: homodyne beam splitter; MC: mode cleaner cavity; PD: photo diode; EOM: electro-optical modulator.}
 \label{setup}
\end{figure}

The light source in our setup, as schematically depicted in Fig.~\ref{setup}, was a high power erbium micro
fiber laser providing about 1.6\,W of continuous-wave radiation at 1550\,nm. The laser beam was first sent
through a ring mode cleaner (MC) cavity with a finesse of 350 and a line width of 1{.}2\,MHz for p-polarized
light. Thus reducing mode distortions of the laser's TEM$_{00}$ spatial mode profile as well as its phase and
amplitude fluctuations at frequencies above the MC linewidth.
Approximately 10mW of the transmitted light served as a local oscillator (LO) for balanced homodyne detection, while the remaining power of about one 1\,W was used for second harmonic generation (SHG) to provide the frequency doubled pump field for the OPA.
Both, SHG and OPA were realized as single-ended standing-wave cavities formed by two mirrors and the non-linear crystal in between. In both cavities we employed a PPKTP crystal of dimension $10\times$2$\times$1\,mm$^3$ with flat, anti-reflection (AR) coated front and end faces. Inside a polyoxymethylene (POM) housing, each crystal is embedded in a copper fixture mounted on a Peltier element. Together with an integrated thermistor this enabled us to actively fine-tune the crystal temperature for efficient nonlinear coupling.
A highly reflective (HR) mirror with a power reflectivity r\,$>$99{.}98\,\% for both the fundamental and second harmonic field faces one AR-side of the crystal and a piezo-driven out-coupling mirror was mounted on the opposite side. The OPA out-coupling mirror had 90\;\% and 20\,\% power reflectivity for 1550\,nm and  775\,nm, respectively. For the SHG  we also used 90\;\% reflectivity for the fundamental but only a marginal reflectivity for the second harmonic. 
The mirrors and the ring-piezo were mounted inside aluminum blocks that were rigidly attached to the POM housing.
Considering the refractive index of $n$=1.816 for PPKTP at 1550\,nm and the spacing of 20\,mm between crystal end faces and mirrors, the cavity waist size $w_0$, free
spectral range FSR, and line width (FWHM) were calculated to be $\omega_0$=60\,$\mu$m, FSR=2.6\,GHz, and FWHM= 43\,MHz, respectively.
When the SHG cavity was locked on resonance it produced up to 800\,mW at 775\,nm, which was separated from the fundamental by a dichroic beam splitter (DBS). The harmonic beam passed a combination of a half waveplate and a polarizing beam splitter for pump power adjustment, a Faraday isolator to prevent the SHG from retro-reflected light, and an electro optical modulator (EOM), and was mode matched to the TEM$_{00}$-mode of another MC cavity (MC$_{775}$) with characteristics equal to those of MC$_{1550}$.
The transmitted beam was then carefully aligned to match the OPA-cavity TEM$_{00}$ mode.
The length control of the cavities in our setup was accomplished by means of a modulation/demodulation (Pound-Drever-Hall, PDH) scheme utilizing custom made EOMs and matched photo detectors. Details on the particular implementation can be found in Fig.~\ref{setup}.
The squeezed states left the OPA in the counter direction to the second-harmonic pump, where another DBS separated the two of them. The measurement of field quadratures variances was
accomplished by means of balanced homodyne detection, for which
the squeezed field was subsequently made to interfere with the
LO on a 50/50-beam splitter. A piezo-actuated steering mirror was employed to shift the LO phase relative to the squeezed field. To adjust the visibility we injected
a control beam through the HR back side of the OPA. This control beam was matched to o the OPA TEM$_{00}$ mode. The light that was transmitted propagated congruent to the mode to be
squeezed, and, by locking the OPA cavity length, could be used to overlap with the LO on the homodyne beam splitter (HBS).
We reached a fringe visibility of 99{.}0\,\%.
The two outputs of the 50/50-beam splitter were each focused down
and detected by a pair of Epitaxx ETX-500 photodiodes. The difference
current was fed to a spectrum analyzer.

\begin{figure}[h]
\includegraphics{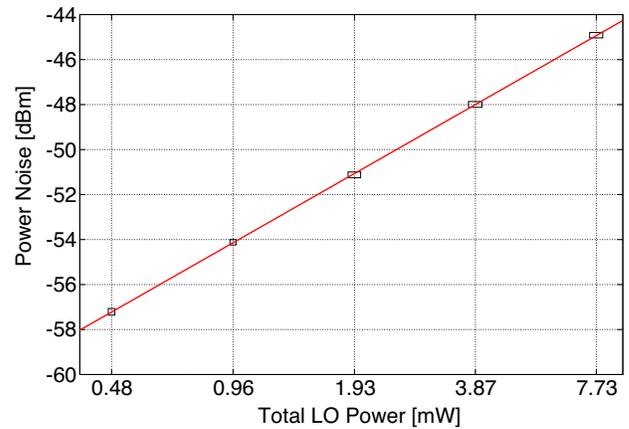}
 \caption{Noise power levels of the homodyne detector were measured at different LO powers at a centre frequency of 5\,MHz with the signal port blocked. Box sizes indicate the standard deviation of the fit and an estimated $\pm$5\% uncertainty of the power meter used. The graph shows that our homodyne detector was quantum noise limited and operated linearly within our measurement regime.} 
 \label{fig1}
\end{figure}

To verify our detector's linearity we took measurements of the vacuum noise power against the incident LO power at a sideband frequency of 5\,MHz as depicted in Fig.~\ref{fig1}. Changing the LO power by a factor of two, entailed a 3\,dB shift of the corresponding noise trace, showing that the detector was quantum noise limited and operated linearly in the measurement regime.

\begin{figure}[h]
\includegraphics{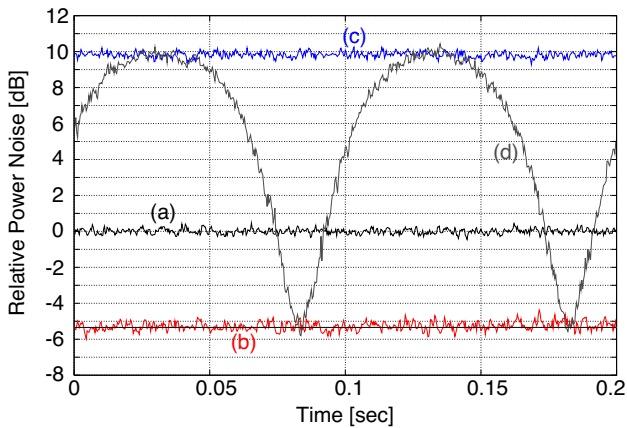}
 \caption{Noise powers of the squeezed light emitted by the OPA at a sideband frequency of 5\,MHz normalized to the shot-noise level (trace (a)).
All traces were recorded with a resolution bandwidth of 300\,kHz and a video bandwidth of 300\,Hz. Squeezing (b) and anti-squeezing (c) curves were averaged twice. Curve (d) was recorded by linearly sweeping the LO-phase which continuously rotated the measured quadrature from anti-squeezing to squeezing.}
 \label{fig2}
\end{figure}

We found the optimum pump power for our OPA to be 300\,mW,
yielding a noise reduction of 5.3\,dB in the squeezed quadrature.
This entailed an increase of 9.8\,dB in the anti-squeezed
quadrature. To switch between the two, a piezo-actuated mirror was
used to phase shift the LO with respect to the squeezed field . The
measured noise curves are depicted in Fig.~\ref{fig2}. Trace (a) is the measured vacuum noise when the signal port of the HBS is blocked. The associated power of the incident LO was approximately 4 \,mW. Upon opening the signal port and injecting the squeezed field of the resonant OPA, trace (d) was recorded by linearly sweeping the LO-phase, thereby changing the measured quadrature from anti-squeezed to squeezed values. By holding the homodyne angle fixed, continuous traces of the squeezing (b) and anti-squeezing (c) were recorded. All traces were recorded at a sideband frequency of 5\,MHz and are, apart from (d), averaged twice. The contribution of electronic dark noise of our detector was negligible (18\,dB below the shot noise) and was not subtracted from the measured data.   

The observed squeezed noise power was 5.3\,dB below shot noise, however, the observed anti-squeezing was about 10\,dB above shot noise, revealing an uncertainty product of about a factor of three above the minimum uncertainty. With an increased pump power we observed further increased anti-squeezing, but a constant squeezing level. Following the argumentation in \cite{Vahlb08} this observation implies that our measurement was not limited by phase noise \cite{Taken07,Franzen06} but by optical losses.
With 0.25\% residual reflectance of our crystal AR coatings and 0.1\%/cm absorption loss within the crystal we  estimate the escape efficiency of the OPA cavity to be 90\%. Together with a propagation loss of approximately 3\%, we estimate the quantum efficiency of our photo detectors to be 90\%$\pm4$\%. We therefore expect that higher levels of squeezing from PPKTP could be observed in future utilizing better photo diodes and an OPA optimized for better escape efficiency.
We note that PPKTP has already been successfully applied for the generation of squeezed and entangled states at wavelengths between 532\,nm and 1064\,nm \cite{Hetet07,Aoki06,Taken07,Goda08NatPhys,Gross08} with the maximum squeezing strength of 9\,dB observed at 860\,nm in \cite{Taken07}. The strongest squeezing to date was reported in~\cite{Vahlb08}  where a MgO:LiNbO$_{3}$ crystal enabled the observation of a noise reduction of 10\,dB below shot noise at 1064\,nm. However, at 1550\,nm the phase matching condition of this material is uncomfortably high and temperature gradients would significantly complicate the stable operation of a squeezed light source. This makes PPKTP the preferable material for the generation of squeezed light at 1550\,nm.

In conclusion, we have demonstrated strong squeezing at the telecommunication wavelength of 1550\,nm. Our experiment proved that PPKTP is an effective material for the generation of squeezed states at this wavelength. The spatio-temporal mode of the squeezed field had a high purity ensuring the compatibility with quantum memories and quantum repeaters. 
By implementing a control scheme according to~\cite{Vahlb06} squeezing in the detection band of current GW detectors can be realized. These detectors are operated at 1064\,nm \cite{Goda08NatPhys}, however, future detector designs might consider silicon as test mass material and the laser wavelength of 1550\,nm in order to the reduce the thermal noise floor.

The authors thank the German Research Foundation and the Centre for Quantum
Engineering and Space-Time Research QUEST for financial support.

%

\end{document}